\documentclass[aps,pra,twocolumn,showpacs]{revtex4}
\usepackage[dvips]{graphicx}
\usepackage{amsmath}
\usepackage{times}
\usepackage{psfrag}

\begin{document}


\title
{
Interference scheme to measure light-induced nonlinearities
in Bose-Einstein condensates
}

\author
{K.~V.~Krutitsky,
K.-P.~Marzlin
and J.~Audretsch
}

\affiliation{
Fachbereich Physik der Universit\"at Konstanz,
Fach M 674,
D-78457 Konstanz, Germany
}

\date{\today}


\begin{abstract}
Light-induced nonlinear terms in the Gross-Pitaevskii equation arise
from the stimulated coherent exchange of photons between two atoms.
For atoms in an optical dipole trap this effect depends on the spatial
profile of the trapping laser beam. Two different laser beams can induce
the same trapping potential but very different nonlinearities. We propose
a scheme to measure light-induced nonlinearities which is based on this
observation.
\end{abstract}

\pacs{03.75.Hh,34.20.Cf,34.80.Qb}

\maketitle

In the framework of the mean-field theory an atomic Bose-Einstein
condensate (BEC) represents a nonlinear system where
the nonlinearity is caused by atomic collisions.
Virtual photon exchange in a BEC of neutral polarizable atoms,
caused by off-resonant optical laser radiation with the Rabi frequency $\Omega$
and the detuning $\Delta$ leads to additional nonlinearities in the
condensate's equation of
motion~\cite{LEN9394,ZHA94a,WAL97,KBA99,DGKA}.
Since the optically induced nonlinearity (OINL) is strongly
suppressed for large detuning, it can be safely neglected for atoms
trapped in optical dipole potentials \cite{diptrap}.
However, for smaller detunings it must be taken into account.
It has been theoretically shown that the OINL in a BEC can lead to nonlinear
diffraction of ultracold atomic beams~\cite{ZHA94a,KBA99},
laser-induced gravity~\cite{DGKA},
and one can create Thirring and gap solitons~\cite{LEN9394},
photonic band gaps and defect states in a periodic BEC~\cite{PBG}.

Inspite of the fact that the OINL can be of the same order of magnitude as
collisional nonlinearities~\cite{KBA99}, the observation of the OINL
is difficult.
Due to spontaneous emission an atomic
BEC in a largely detuned laser field typically decoheres at a rate
of
$
\gamma_{\mbox{\scriptsize dec}}
:=
\gamma | \Omega|^2/\Delta^2
$,
where $\gamma$ is the spontaneous emission rate of a single atom in free space
(see, for instance, Ref.~\cite{kpm01}). Hence,
in order to avoid spontaneous emission one has to consider only
short interaction times $T$ such that
$T\gamma_{\mbox{\scriptsize dec}} \ll 1$.
In this case, the effect of the OINL essentially amounts to a small
nonlinear phase imprint on the BEC.
The goal of this paper is to develop a scheme to measure this phase imprint.

We would like to note that decoherence of another type can be caused by the fact
that there is a distance (the so-called Condon distance) at which a pair of atoms is
resonant with the light, even if the light is detuned from the atomic resonance
for a single atom~\cite{WBZJ99}. However, we neglect this type of decoherence, because in
the case of $^{87}$Rb at the typical density of $10^{14}$ cm$^{-3}$
and the detuning $\Delta \sim 1$ GHz, the Condon distance
is almost five times smaller than the mean interatomic distance.

The detailed derivation of the OINL in a BEC
was given in Ref.~\cite{KBA99}. Its features are
determined by the properties of the local electric field,
which governs the evolution of the condensate's wave function $\psi$
in the field of electromagnetic radiation. In the case of
low light intensity and in the slowly varying (in time) amplitude
approximation, we have the following equation for the local electric-field
strength~\cite{KBA99,MOR95}
\begin{eqnarray}
\label{e-local}
&&
{\bf E}_{loc}^+({\bf x})
=
{\bf E}_{in}^+({\bf x})
+
\int
d{\bf x}'
e^{
    i k_L R
  }
{\bf P}^+({\bf x}')
\\
&&
\times
\left[
    \frac{3\cos^2\theta-1}{R^3}
    -
    i k_L
    \frac{3\cos^2\theta-1}{R^2}
    -
    k_L^2
    \frac{\cos^2\theta-1}{R}
\right]
\nonumber
\;,
\end{eqnarray}
where
$R=|{\bf x}-{\bf x}'|$,
$
{\bf P}^+({\bf x})
=
\alpha
\left|
    \psi({\bf x})
\right|^2
{\bf E}_{loc}^+({\bf x})
$
the medium polarization,
$\alpha=-d^2/(\hbar\Delta)$ the atomic polarizability,
and $\theta$ the angle between the vector ${\bf R}$ and the dipole axis.
${\bf E}_{in}^+$ in Eq.(\ref{e-local}) is an incident
laser field with the frequency $\omega_L=c k_L$, which is close to the
frequency $\omega_0$ of the electric-dipole transition.
The integral term describes {\it dynamical dipole-dipole interaction}
and leads to the OINL in the dynamics of
the condensate's wave function $\psi$.
As it was shown in Ref.~\cite{KBA99}, Eq.(\ref{e-local}) describes
subsequent pair interactions between the atoms when an off-resonant laser photon enters
the medium of ground-state atoms and {\it virtually} excites an atom, then this
atom goes back to the ground state and emits a photon, which is absorbed by
another atom and so on.
We would like to stress that the mechanism described above is different from
the (quasi)photoassociation, when a pair of atoms first absorbs a photon and
undergoes a virtual transition to an electronically excited quasimolecular state
and then it reemits the photon and returns to the initial electronic state~\cite{FKSW96}.

The dipole field in Eq.(\ref{e-local}) contains three terms, which have different
dependence on $R$. The first one ($\sim 1/R^3$) is a near field
and gives a significant contribution at distances much smaller than
the laser radiation wavelength. The last one ($\sim 1/R$) is a far
field and gives a dominant
contribution at distances greater than the wavelength. At distances
of the order of wavelength all the three terms are of
the same order of magnitude.
They must be taken into account for condensates with a typical
size of several optical wavelengths.

The near and far fields have different dependencies on the angle $\theta$.
The sign of the near-field term can be either negative or positive depending
on $\theta$ and the near field vanishes after the angular integration
in the case of a homogeneous density distribution.
The sign of the far field does not depend on $\theta$ and it does
not vanish for any atomic distribution.
This leads to an attractive effective nonlinear potential in a large
system of interacting optically induced dipoles.

\begin{figure}[ht]
\centering
  \includegraphics[width=5cm]{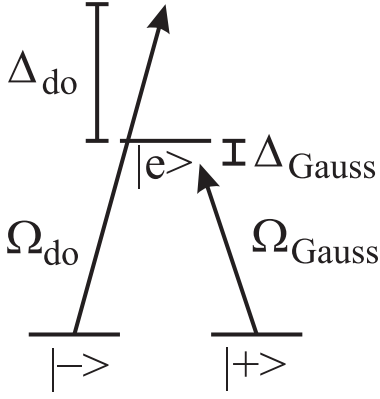}
\caption{
         Level scheme of an atomic spin-1 BEC interacting with off-resonant
 circularly polarized laser waves.
 }
\label{lambda}

\end{figure}

The above discussion refers to the microscopic electrodynamics of a BEC.
Since usually the condensate density does not change significantly
over one wavelength, the nonlinear optical potential can also be described
in terms of a macroscopic (averaged) electric field~\cite{KBA99}.
This approach is equivalent to the microscopic one
and allows to simplify the equation for the condensate wave function.
As a result of the macroscopic approach the OINL can be written as
a local attractive term [see Eq.(\ref{kappa})] in the Gross-Pitaevskii equation (GPE).

In this paper, we are concerned with atoms possessing two internal
ground states characterized by the magnetic
quantum number $m=\pm 1$. In the presence of two circularly polarized
laser beams and in the case of
a dilute gas with the level configuration shown in Fig.~\ref{lambda}
the two collective wave functions  $\psi_\pm$  fulfill
the modified GPE \cite{KBA99}
\begin{eqnarray}
\label{gpm}
&&
  i \hbar
  \frac{\partial \psi_\pm}{\partial t}
  =
  \bigg \{
  -\frac{\hbar^2 \nabla^2}{2M}
  +
  V_{\mbox{\scriptsize opt},\pm}({\bf x})
  \\
  &&
  +
  \left[
      \kappa_s + \kappa_a
      + \kappa_{\mbox{\scriptsize opt},\pm}({\bf x})
  \right]
  \left|
      \psi_\pm
  \right|^2 +
  \left( \kappa_s - \kappa_a \right)
  \left| \psi_\mp \right|^2
  \bigg \}
  \psi_\pm
\; ,
\nonumber
\end{eqnarray}
where $M$ is the atomic mass.
It is assumed that the detunings $\Delta_\pm$ of the laser frequencies
from the respective atomic transitions satisfy the inequality
$|\Delta_-| \gg |\Delta_+|$, such that there is no
population exchange through Raman transitions
between the components and the condensate is similar to
a two-species one \cite{KBA99}. The strength
of the collision-induced nonlinearity is described by the coefficients
$\kappa_{s,a} := 4 \pi \hbar^2 a_{s,a}/M$, where
$a_{s,a}$ are symmetric and antisymmetric scattering lengths,
respectively. For large detunings ($\left|\Delta_\pm\right|\gg\gamma$)
one can neglect the refractive index of the gas even for interatomic distances
of the order of the transition wavelength $\lambda_0$.
In such situation the Rabi frequencies $\Omega_\pm({\bf x})$, which are proportional
to the electric-field strength of the $\sigma_\mp$-polarized
laser beams, do not depend on the atomic density and the potential
\begin{equation}
  V_{\mbox{\scriptsize opt},\pm}({\bf x})
  =
  \frac{\hbar \left| \Omega_\pm({\bf x}) \right|^2}{\Delta_\pm}
\label{optpot}
\end{equation}
does not lead to any nonlinearity in Eq.(\ref{gpm}).
The OINL coefficient is also, in general, position dependent and in the
case of low light intensity
($\left|\Omega_\pm\right|^2 \ll \Delta_\pm^2 $) is given by
\begin{equation}
  \kappa_{\mbox{\scriptsize opt},\pm}({\bf x})
  =
  -2\pi
  \hbar
  \gamma
  \frac{\left| \Omega_\pm({\bf x}) \right|^2}{\Delta_\pm^2}
  \left(\frac{\lambda_0}{2\pi}\right)^3
\label{kappa}
\;.
\end{equation}

Our scheme to measure the OINL is based on the observation that
$V_{\mbox{\scriptsize opt},\pm}({\bf x})$ can induce about the same
mechanical forces on atoms even for very different laser beam
profiles.
A doughnut laser beam with Rabi frequency
$
  \Omega_{\mbox{\scriptsize do}}({\bf x})
  =
  \Omega^0_{\mbox{\scriptsize do}}
  \exp(-r_\perp^2 /w^2)
  \exp(i k_L z)
  (x+i y)/w
$
and a Gaussian laser beam with
$
  \Omega_{\mbox{\scriptsize Gauss}}({\bf x})
  =
  \Omega^0_{\mbox{\scriptsize Gauss}}
  \exp(-r_\perp^2 /w^2) \exp(i k_L z)
$
produce the optical potentials
\begin{eqnarray}
  V_{\mbox{\scriptsize do}}
  &=&
  \hbar
    \frac
    {|\Omega^0_{\mbox{\scriptsize do}}|^2 }
    {\Delta_{\mbox{\scriptsize do}}}
    \frac{r_\perp^2}{w^2}
    \exp(-2 r_\perp^2 /w^2)
  \;,
\label{potdo}
\\
  V_{\mbox{\scriptsize Gauss}}
  &=&
  \hbar
    \frac
    {|\Omega^0_{\mbox{\scriptsize Gauss}}|^2 }
    {\Delta_{\mbox{\scriptsize Gauss}}}
     \exp(-2 r_\perp^2 /w^2)
  \;,
\label{potGauss}
\end{eqnarray}
$r_\perp := \sqrt{x^2+y^2}$.
Thus, apart from a constant energy shift
$
\delta_V
=
\hbar |\Omega^0_{\mbox{\scriptsize Gauss}}|^2 /
\Delta_{\mbox{\scriptsize Gauss}}
$,
both beams provide approximately the same harmonic trapping
potential with trap frequency $\omega_\perp$ provided that
\begin{equation}
  \frac{M \omega_\perp^2}{2}
  =
  \hbar
    \frac
    { |\Omega^0_{\mbox{\scriptsize do}}|^2 }
    { w^2 \Delta_{\mbox{\scriptsize do}} }
    =
    -2\hbar
    \frac{|\Omega^0_{\mbox{\scriptsize Gauss}}|^2 }{w^2
          \Delta_{\mbox{\scriptsize Gauss}}}
\label{rabicond}\end{equation}
holds. Note that $\Delta_{\mbox{\scriptsize do}}$ must be positive and
$\Delta_{\mbox{\scriptsize Gauss}}$
negative in order to provide a trapping potential.

A very important physical difference between these two potentials
is that in the doughnut potential atoms are low-field seekers,
whereas in the Gaussian beam they tend to accumulate
at positions with high laser intensity.
This implies that the atoms experience a strong
OINL in a Gaussian beam, but a negligible
nonlinearity in a doughnut beam.
This effect is the fundamental phenomenon on which our proposal is based on.

We would like to emphasize that in this situation any difference
in the evolution of the two BEC components $|\pm \rangle$
would be produced by the OINL alone.
In the harmonic approximation, the trapping potentials only differ by
an unimportant spatially homogeneous value $\delta_V$.
Furthermore, for $^{87}$Rb the antisymmetric scattering length $a_a$
is negligible (see numerical values below),
so that the collision-induced nonlinearities are
identical for both components.

The measurement scheme consists of the following steps.

(i) {\it $\pi/2$ Raman pulse preparing a superposition of the
states $|+\rangle$ and $|-\rangle$}.
We assume that the BEC is
initially prepared in the state $|-\rangle$
with a spatial wave function $\psi({\bf x},0)$
corresponding to the stationary collective ground state
in a doughnut beam with $\sigma_+$ polarization
($\Omega_- = \Omega_{\mbox{\scriptsize do}}$).
The detuning of the doughnut beam is
huge so that both the spontaneous emission and the OINL can be neglected.
Half of the atoms are then transferred to the state $|+\rangle$
by a $\pi/2$ laser pulse. Since powerful experimental
techniques such as stimulated Raman adiabatic passage~\cite{STIRAP} do exist,
we can assume that the
transfer is nearly perfect and can be described by the
unitary operator
\begin{equation}
\label{U}
U
=
\frac{e^{-i\varphi_s}|-\rangle + |+\rangle}{\sqrt{2}}
\langle -|
+
\frac{e^{i\varphi_s}|+\rangle - |-\rangle}{\sqrt{2}}
\langle +|
\;,
\end{equation}
where $\varphi_s$ is the phase of the $\sigma_-$-polarized Stokes laser.
It can be freely chosen and will be discussed later.
After this transformation with the duration $T_{\pi/2}$,
the BEC is in a coherent superposition
of the states $|\pm \rangle$ with the wave functions
\begin{eqnarray}
\psi_-({\bf x},T_{\pi/2})
&=&
\frac{\psi({\bf x},0)}{\sqrt{2}}
\exp(-i\varphi_s)
\;,
\nonumber\\
\psi_+({\bf x},T_{\pi/2})
&=&
\frac{\psi({\bf x},0)}{\sqrt{2}}
\;.
\label{init-cond}
\end{eqnarray}

(ii) {\it Phase imprint.}
In the second step, we apply an additional $\sigma_-$ polarized
Gaussian laser beam to the BEC so that
$\Omega_+ = \Omega_{\mbox{\scriptsize Gauss}}$.
Both detuning and intensity of this beam are much smaller
than that of the doughnut beam and are chosen so that relation
(\ref{rabicond}) holds. For small enough detuning the OINL is
not negligible for this beam but completely negligible for the doughnut
beam. Thus, only the wavefunction for atoms in state $|+\rangle$
is changed by the OINL.
Since all other parameters are held constant
this is the only change the system experiences beside the phase shift
caused by the potential difference $\delta_V$.

Since the interaction time $T$ must be small as compared to
$1/\gamma_{\mbox{\scriptsize dec}}$
it has to be of the order of $10\ \mu$s in a realistic experiment.
For such a short interaction time, the spatial distribution of rubidium atoms
does not change much and, therefore,
it is possible to neglect the center-of-mass motion.
The wave functions at time $t=T+T_{\pi/2}$ then can be written in the form
\begin{eqnarray}
\label{AFI}
\psi_-({\bf x},t)
&=&
\frac{\psi({\bf x},0)}{\sqrt{2}}e^{-i\mu T/\hbar}
e^{-i\varphi_s}
\;,
\\
\psi_+({\bf x},t)
&=&
\frac{\psi({\bf x},0)}{\sqrt{2}}e^{-i\mu T/\hbar}
e^{i(\varphi_V+\varphi_{\mbox{\scriptsize OINL}})}
\;,\nonumber
\end{eqnarray}
where
$ \varphi_{\mbox{\scriptsize OINL}}= -
  \kappa_{\mbox{\scriptsize opt},+}({\bf x})
  \left| \psi_+ \right|^2  T/\hbar $
is the phase shift produced by the OINL, i.e., the quantity we
are interested in. $ \mu $ is the chemical potential and
$\varphi_V := \delta_V T/\hbar $.

(iii)  {\it Second $\pi/2$ Raman pulse}.
After the phase imprint the Gaussian laser beam is switched off.
The interferometric scheme is completed by applying a second
$\pi/2$ Raman pulse that is identical to the first
one. After this pulse a number of
\begin{equation}
\label{N}
  N_-  = \int |\psi({\bf x},0)|^2
  \sin^2 \frac{\varphi_{\mbox{\scriptsize OINL}}+
   2 \varphi_{s}+ \varphi_V}{2}
  \,d{\bf x}
\end{equation}
atoms are left in the state $|-\rangle$, where they are trapped by
the doughnut beam. The atoms in the state $|+\rangle$ are not trapped
anymore and are dispersed in a time that typically is of
the order of a few milliseconds.
To make $N_-$ a signal of the OINL alone, one has
to eliminate the influence of the potential difference $\delta_V$
which can be done by choosing $\varphi_s = - \varphi_V /2$. In this
case, $N_-$ vanishes in the absence of the OINL.

\begin{figure}[t]
\centering
   \centering
   \hspace{-1.5cm}
      \includegraphics[width=5.5cm]{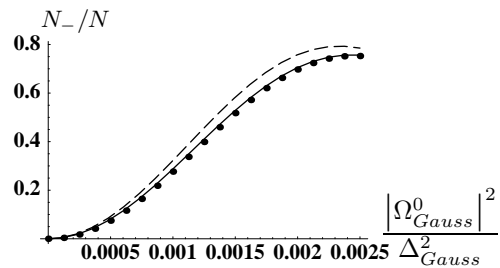}
\caption{
          Number of atoms in the trapped state $|-\rangle$ after the
          interaction with the Raman pulses.
          Dots -- full numerical simulation,
          solid line -- numerical integration of Eq.(\ref{N}) with
          exact ground state,
          dashed line -- analytical
          calculations based on the Thomas-Fermi approximation
          [Eq.(\ref{N-TFA})].
        }
\label{rest}
\end{figure}

In order to get concrete estimations,
we have considered a BEC of $N=10^5$ rubidium atoms
($^{87}$Rb, $M = 1.45 \times 10^{-25}\ kg$, $a_s=5.4\ nm$,
$a_a=-0.05\ nm$~\cite{Ho},
$\lambda_0=780$ nm, $\gamma=38$ MHz) in a
trapping potential of the form (\ref{potdo})
and with the parameters~\cite{BBDHAES}
$\Delta_{\mbox{\scriptsize do}} = 1.1 \times 10^{15}$ Hz,
$\Omega^0_{\mbox{\scriptsize do}} = 3.15 \times 10^{10}$ Hz
(which corresponds to the oscillation frequency
$\omega_\perp = 2\pi \times 576$ Hz),
$w = 10\ {\mu}$m. Using these parameters,
we have performed analytical two-dimensional (2D) calculations along the lines described
above by employing the Thomas-Fermi approximation for the calculation of
the ground state wave function $\psi({\bf x},0)$.
The condensate has been assumed to be homogeneously extended along the
$z$ direction with a length of $L_z=20\ {\mu}m$,
which gives the following value of the
Thomas-Fermi radius:
$ R_{TF} = 2 \left( \hbar/M\omega_\perp \right)^{1/2}
 \left[ (a_s+a_a)N/L_z \right]^{1/4}
 = 2\ {\mu}m
$.
The number of
atoms remaining in the trapped state $N_-$ can be worked out according
to Eq.(\ref{N}), assuming that the condensate size is much smaller than
the laser width $w$. The result is given by
\begin{eqnarray}
\label{N-TFA}
&&
N_-
=
N
\left(
    \frac{1}{2}
    -
    \frac{\sin\varepsilon}{\varepsilon}
    -
    \frac{\cos\varepsilon-1}{\varepsilon^2}
\right)
\;,
\\
&&
\varepsilon
=
\frac
{|\Omega^0_{\mbox{\scriptsize Gauss}}|^2 }
{\Delta^2_{\mbox{\scriptsize Gauss}}}
\gamma T
\left(
    \frac{\lambda_0}{2\pi}
\right)^3
\frac{M\omega_\perp}{2\hbar}
\sqrt{\frac{N}{\left(a_s+a_a\right)L_z}}
,
\nonumber
\end{eqnarray}
and is shown in Fig.~\ref{rest} (dashed line) for the interaction
time $T=10\ {\mu}s$.
As one clearly sees, the number $N_-$
of trapped atoms increases with the increase
of the light-induced nonlinearity which is proportional to
$
\left| \Omega^0_{\mbox{\scriptsize Gauss}}\right|^2 /
\Delta_{\mbox{\scriptsize Gauss}}^2
$
and reaches about $0.75 N$ at
$
\left| \Omega^0_{\mbox{\scriptsize Gauss}}\right|^2 /
\Delta_{\mbox{\scriptsize Gauss}}^2
=0.0025
$.
However, to fulfill $T \gamma_{\mbox{\scriptsize dec}} \ll 1$
the parameter
$
\left| \Omega^0_{\mbox{\scriptsize Gauss}}\right|^2 /
\Delta_{\mbox{\scriptsize Gauss}}^2
$
must not be higher than about $0.001$, i.e., values of
$N_-$ up to $0.25 N$ could be observable.

\psfrag{N}{$N_-/N$}
\psfrag{p}{\Large{$\frac{\left|\Omega^0_{Gauss}\right|^2}{\Delta_{Gauss}^2}$}}

To confirm our analytical results, we have also
performed 2D numerical calculations
in which the center-of-mass motion, the full form of the potentials
$V_{\mbox{\scriptsize opt},\pm}$, and all nonlinearities
are taken into account.
The initial ground-state wave function $\psi$ has been obtained by
numerical solution of the two-dimensional Gross-Pitaevskii equation.
For the time evolution, we have
employed the split-step method (see, e.g., Ref.~\cite{deVries86})
and combined it with the
imaginary time propagation technique (see, e.g.,
Ref.~\cite{dalfovo96})
to find the ground state.

Starting from state (\ref{init-cond}), we have solved the time evolution
of the BEC which is governed by Eq.(\ref{gpm}).
After that the transformation $U$ of the two-component wave function
has been performed and the number of atoms in the trapped state $N_-$
has been worked out. The dependence of $N_-$ on the parameter
$\left| \Omega^0_{\mbox{\scriptsize Gauss}}\right|^2 /
\Delta_{\mbox{\scriptsize Gauss}}^2$ is shown in
Fig.\ref{rest} (dots). The results are in good
agreement with the approximate analytical calculations. However,
the Thomas-Fermi approximation leads to somewhat higher values of
$N_-$ compared to the exact numerical calculations, because in
neglecting the center-of-mass motion
the Thomas-Fermi approximation gives a slightly higher central density.
The comparison of our analytical and
numerical calculations shows that the center-of-mass motion does not
influence significantly $N_-$, which is mainly determined by
the form of the initial atomic distribution and the parameters
of the laser radiation in the middle of the trap.
After the second Raman pulse those atoms transferred to state
$|+ \rangle$ are dispersed within about 1 ms,
because the Gaussian laser is switched off. Since the transfer is not adiabatic
the number of atoms in the state $|- \rangle$ is reduced, but the shape of the
wave function remains almost the same. If there would be no atomic collisions
this would be again a stationary state of the corresponding Schr\"odinger equation,
but in the presence of collisions this is not the case and the remaining cloud
of atoms in state $|- \rangle$ starts to oscillate radially. They
can be detected by taking an absorbtion image with resonant laser light.

In the present paper, we have developed an interference scheme that
allows to measure the OINL caused by dynamical dipole-dipole interaction
in the field of an optical laser.
The observation of this effect
would help to improve our understanding of the interactions between
light and atoms.

We thank Markus Oberthaler for very stimulating discussions.
This work has been supported by the Deutsche For\-schungsgemeinschaft
and the Optikzentrum Konstanz. K.V.K. would also like to
thank the Alexander-von-Humboldt Stiftung for financial support.


\end{document}